# High-angle diffraction of a Gaussian beam by the grating with embedded phase singularity

A. Bekshaev[1]*, O. Orlinska[1], M. Vasnetsov[2]

[1]*I.I. Mechnikov National University, Dvorianska 2, 65082, Odessa, Ukraine*
[2]*Institute of Physics, National Academy of Sciences, Prospect Nauki 46, Kiev, 03028 Ukraine*

**Abstract**

Spatial characteristics of the optical-vortex (OV) beams created during the Gaussian beam diffraction by a grating with groove bifurcation are analyzed theoretically and numerically. In contrast to previous works, condition of small-angle diffraction is no longer required and the diffracted beam can be strongly deformed. This causes the intensity profile rotation and the high-order OV decomposition into a set of secondary single-charged OVs. These effects are studied quantitatively and confronted with similar properties of a Laguerre-Gaussian beam that undergoes astigmatic telescopic transformation. In contrast to the latter case, the secondary OVs do not lie on a single straight line within the beam cross section, and morphology parameters of the individual secondary OVs carried by the same beam are, in general, different. Conditions for maximum relative separation of the secondary OVs with respect to the beam transverse size are specified. The results can be used for practical generation of OV beams and OV arrays with prescribed properties.

*PACS*: 42.25.Bs; 42.25.Fx; 42.40.Jv; 42.50.Tx; 42.60.Jf; 42.90.+m

*Keywords*: Optical vortex; Generation; Computer-synthesized hologram; High-angle diffraction; Spatial structure; Beam evolution

*Corresponding author. Tel.: +38 048 723 80 75
*E-mail address*: bekshaev@onu.edu.ua (A.Ya. Bekshaev)

## 1. Introduction

Holographic methods are among the most suitable and universal means to obtain optical beams with predicted special structure, and the optical vortices (OV), i.e. beams with helical wavefront shape [1–5], are not exclusion. Usually the OVs are produced due to diffraction of a regular wave with smooth wavefront (incident beam) on a special computer-generated hologram (CGH) that represents a sort of diffraction grating with a groove bifurcation forming the so called "fork" structure (see Fig. 1) [6–10]. If a single groove divides into $m + 1$ branches (in Fig. 1 $m = 1$), the $n$-order diffracted beam acquires the OV with topological charge

$$l = mn. \qquad (1)$$

Integer number $m$ is usually referred to as the topological charge of the phase singularity "embedded" in the CGH [11–13]; both $m$ and $n$ can be positive or negative.

Properties of the diffracted beams carrying the OVs created in this process essentially depend on many conditions, determining the diffraction regime: relative disposition of the CGH and the incident beam, diffraction order, spatial frequency of the CGH, etc. In many applications it is necessary to generate OV beams with prescribed properties or, at least, to predict characteristics of an OV obtained under certain conditions. To this purpose, detailed studies of the process of OV generation in a CGH with the "fork" structure have been undertaken in recent years [9–15]. In these works, considerable successes were achieved in theoretical and experimental investigation of spatial properties of the OV beams produced by the "fork" CGH in the nominal (the incident beam is Gaussian with axis orthogonal to the grating plane and passing exactly through the bifurcation point) and misaligned (the incident beam axis is inclined and/or shifted with respect to the nominal position) configuration. However, almost all known results were found for the case of small diffraction angle $\theta \ll 1$ (see Fig. 1). This condition implies that the OV-producing CGH has low spatial frequency (below 100 grooves per millimeter) and only low-order diffracted beams are admissible.

At the same time, the high-angle diffraction arrangement is rather attractive for the search of new possibilities to create the OV beams with unusual properties. Besides, such situations can be advantageous in view of the diffraction efficiency and concentration of the diffracted energy in a desirable diffraction order. That is why the knowledge of properties and behavior of OV beams obtained in the "fork" CGH under conditions of high-angle diffraction is desirable and constitutes a relevant problem.

In this paper, we present the results of theoretical consideration of this problem which is based on the general mathematical model developed in Ref. [12]. Geometrical configuration of the model is presented in Fig. 1. The CGH is considered as a planar transparency with spatially inhomogeneous transmittance $T(\mathbf{r}_a)$ where $\mathbf{r}_a = (x_a, y_a) = (r_a\cos\phi, r_a\sin\phi)$ is the radius-vector; the coordinate frame is chosen so that its origin coincides with the bifurcation point and axis $y_a$ is parallel to the grating grooves far from the "fork" (see Fig. 1). The nominal axis of a readout (incident) beam coincides with axis $z_a$ forming a 3D Cartesian frame with axes $x_a$ and $y_a$ (at the grating plane $z_a = 0$).

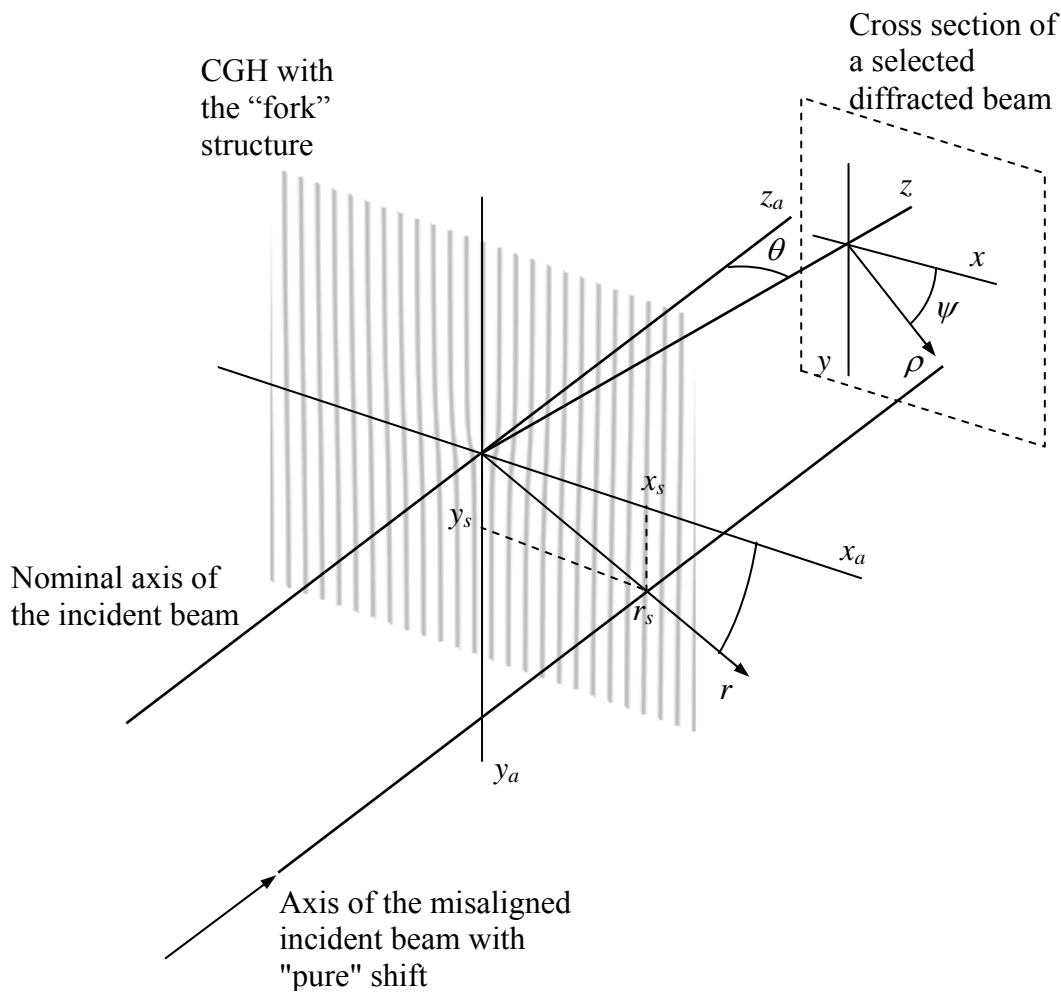

Fig. 1. Geometrical conditions of the beam transformation in a CGH.

Behind the grating, due to periodicity of the transmittance function $T(\mathbf{r}_a)$, a set of paraxial beams (diffraction orders) is formed, propagating in directions specified by condition

$$\sin\theta = \frac{2\pi n}{kd} \qquad (2)$$

where $d$ is the grating period and $k$ is the wave number of the monochromatic incident radiation [16]. To describe the field of a separate diffraction order, it is suitable to introduce the associated coordinate frame $(x, y, z)$ with the origin in the bifurcation point (in Fig. 1, axes $x$ and $y$ are translated along axis $z$ to the current cross section). As was shown before [11,12], a diffracted beam of a separate order propagates along its axis $z$ as a paraxial beam and its field can therefore be represented as $E(x,y,z) = u(x,y,z)\exp(ikz)$ with the slowly varying complex amplitude $u(x,y,z)$ [17]. The spatial distribution of the complex amplitude of $n$-th diffraction order can be determined by equation [12]

$$u_l(x,y,z) = \frac{k}{2\pi i z}\cos^2\frac{\theta}{2} \cdot \int u_a(x_a, y_a) e^{il\phi} \exp\left\{\frac{ik}{2z}\left[(x - x_a\cos\theta)^2 + (y - y_a)^2\right]\right\} dx_a dy_a \quad (3)$$

where $l$ is defined by Eq. (1) and $u_a(x_a, y_a)$ is the complex amplitude distribution of the incident beam at $z_a = 0$. In fact, Eq. (3) describes propagation of the paraxial beam with initial complex amplitude distribution $u_a(x_a, y_a)\exp(il\phi)$, "squeezed" along axis $x$ proportionally to the squeezing coefficient

$$\sigma = (\cos\theta)^{-1}. \quad (4)$$

Eq. (3) was the basis of previous works [11–13,15] but there, due to small-angle geometry, condition $\cos\theta \approx 1$ was accepted. Now this limitation is removed.

However we still restrict our consideration by the case of incident Gaussian beam possibly deviating from the nominal configuration. In this case, a situation when the incident beam is inclined and translated with respect to the nominal axis can formally be treated as a "pure" translation of the incident beam parallel to axis $z_a$ [13]. So without loss of generality we can analyze transformation of the incident beam with complex amplitude distribution

$$u_a(x_a, y_a) = \exp\left[-\frac{(x_a - x_s)^2 + (y_a - y_s)^2}{2b^2} + ik\frac{(x_a - x_s)^2 + (y_a - y_s)^2}{2R}\right] \quad (5)$$

where $x_s$, $y_s$ are Cartesian components of the incident beam translation, $b$ and $R$ are the beam transverse size and the wavefront curvature radius. Further simplification is provided by the fact that the situation with arbitrary wavefront curvature can be reduced to the case of a plane wavefront. For small-angle diffraction, this was proven in Ref. [13]; in general, after (5) is substituted into Eq. (3), the result can be presented in the form

$$u_l(x,y,z) = \cos^2\frac{\theta}{2} \cdot \exp\left\{\frac{ik}{2R}\left[\frac{(x - x_a\cos\theta)^2}{\cos^2\theta + \frac{z}{R}} + \frac{(y - y_a)^2}{1 + \frac{z}{R}}\right]\right\} \cdot u_{le}(x_e, y_e, z_e, \theta_e) \quad (6)$$

where

$$x_e = \frac{x\cos\theta + x_s \frac{z}{R}}{\sqrt{\cos^2\theta + \frac{z}{R}}\sqrt{1+\frac{z}{R}}}, \quad y_e = \frac{y + y_s \frac{z}{R}}{1+\frac{z}{R}}, \quad z_e = \frac{z}{1+\frac{z}{R}}, \quad \cos\theta_e = \frac{\sqrt{\cos^2\theta + \frac{z}{R}}}{\sqrt{1+\frac{z}{R}}}, \quad (7)$$

and function

$$u_{le}(x,y,z,\theta) = \frac{k}{2\pi i z}$$

$$\times \int \exp\left[-\frac{(x_a - x_s)^2 + (y_a - y_s)^2}{2b^2}\right] e^{il\phi} \exp\left\{\frac{ik}{2z}\left[(x - x_a\cos\theta)^2 + (y - y_a)^2\right]\right\} dx_a dy_a \quad (8)$$

describes transformation of the incident Gaussian beam that intersects the CGH exactly at its waist. Note that allowance for the non-planar wavefront is not just scaling and shifting of the coordinates, as in Ref. [13], especially due to dependence of the effective diffraction angle $\theta_e$ on $z$ (see third Eq. (7)). Nevertheless, Eqs. (6) and (7) reduce the problem to analysis of Eq. (8), i.e. enable to restrict our study by incident beams with planar wavefront ($R = \infty$).

It is convenient to introduce the scaled dimensionless parameters

$$\xi_j = x_j/b, \quad \eta_j = y_j/b \quad (9)$$

($j$ – arbitrary index, or no index) and

$$\zeta = z/z_R \quad (10)$$

where

$$z_R = kb^2 \quad (11)$$

is the Raleigh range of the incident Gaussian beam [17]. Then Eq. (8) reduces to the dimensionless form which will be used in further analysis

$$u_{le}(\xi,\eta,\zeta,\theta) = \frac{1}{2\pi i \zeta} \int u_a(\xi_a,\eta_a) e^{il\phi} \exp\left\{\frac{i}{2\zeta}\left[(\xi - \xi_a\cos\theta)^2 + (\eta - \eta_a)^2\right]\right\} d\xi_a d\eta_a \quad (12)$$

where

$$u_a(\xi_a,\eta_a) = \exp\left[-\frac{(\xi_a - \xi_s)^2 + (\eta_a - \eta_s)^2}{2}\right]. \quad (13)$$

Although Eq. (12) is derived for Gaussian beams when Eq. (13) holds, generally it is valid for arbitrary incident beams provided that there exists a certain characteristic transverse scale $b$ of the incident beam profile enabling to introduce the dimensionless beam parameters analogously to (9) – (11).

## 2. General description of spatial profile of the generated OV beams

In this paper we examine the diffracted beam properties by means of numerical evaluation of Eq. (12), addressing when possible to approximate analytical estimates derived in the Appendix. Numerical calculations are made for monochromatic incident beams of a He-Ne laser with the wavelength $\lambda = 0.6328$ μm ($k = 9.93 \cdot 10^4$ cm). In order to concentrate the attention on effects associated with the high-angle diffraction, in the numerical analysis we exclude "misaligned" situations when the incident beam axis is displaced from the bifurcation point. Therefore, keeping Eqs. (5) – (8), (13) with non-zero incident beam shift for further references, in the examples of this paper we will employ their simplifications following from the assumption $x_s = y_s = 0$. In such conditions, at any cross section the beam transverse pattern obeys the central symmetry with respect to the beam axis: this follows from the symmetry of the transformation scheme of Fig. 1 reflected by the symmetry of Eq. (3) and is confirmed by Fig. 2 which represents the beam profiles seen from the positive end of axis $z$ (against the beam propagation).

The specific feature of high-angle OV generation is that in this case the output beam, simultaneously with its formation, experiences transverse deformation: it is squeezed in the plane of the beam deflection (horizontal plane in Fig. 1), which is expressed by term $\cos\theta$ in the integrand exponents of Eqs. (3) and (8) or (12). This deformation corresponds to the astigmatic telescopic transformation of the obtained OV beam with the squeezing coefficient (4).

The simplest situation of this sort occurs upon generation of a single-charged OV beam (see Fig. 2, 1st row). The output beam evolution shows distinct features that are common with recently discussed astigmatic telescopic transformation of the Laguerre-Gaussian (LG) beams [18]. Having broken circular symmetry from the very beginning of its formation, the formed diffracted beam undergoes general rules of evolution of OV beams with symmetry breakdown [5,13,18–21]. The hidden "vortex"-type energy circulation partly transforms into the "asymmetry" circulation that is seen by the visual rotation of the transverse beam profile upon propagation; the rotation agrees with the energy circulation in the "prototype" circularly symmetric OV beam. If the near-field beam profile is elongated vertically (as it happens in the scheme of Fig. 1 and is shown in the first and second columns of Fig. 2), during propagation it is transformed to the horizontally elongated shape. This is a well-known common feature of asymmetrically deformed beams, for example, of an asymmetric Gaussian beam whose evolution is presented in the second row of Fig. 2 for comparison. However, the manners in which the beam shape evolves from the near-field to the far-field "limits" look quite differently. The Gaussian beam profile gradually modifies due to higher rate of the diffraction expansion in the horizontal direction, passing all the intermediate configurations with the initial rectangular symmetry (vertical ellipse → circle → horizontal ellipse). To the contrary, evolution of the OV beam rather looks as a sort of rotation, or "tumble"; the initial rectangular symmetry "temporary" disappears in all the intermediate cross sections and is restored only in the far field.

Also, Fig. 2 demonstrates some peculiar properties of the CGH-generated beams distinguishing them from their LG analogs. The most impressive is that at early stages of the beam evolution a "ripple" structure modulates the complex amplitude distribution, which is plainly visible in the first and second columns of Fig. 2. The nature of the ripples was studied in works discussing the small-angle diffraction in the CGH [12,13] and lies in the following. The bifurcation point is a singular point of the CGH-modeling transparency analogous to the axial discontinuity of a spiral phase plate, also employed to generate OV beams [22,23]. When the incident beam passes the CGH, this singular point serves a source of the divergent spherical wave that copropagates with the "regular" diffracted OV beam and interferes with it. At the distance $z$ behind the CGH, the interference pattern is similar to that formed by a spherical wave with the wavefront curvature radius $z$ and a plane wave: the $p$-th fringe is separated from the beam axis by approximately $r_p = \sqrt{p\lambda z}$, or, in dimensionless units (9), (10), $\rho_p = \sqrt{2\pi p \zeta}$. To be visible, the fringe should appear within the beam bright spot, i.e. the condition should be fulfilled $\rho_p \lesssim 1$, or $\zeta \lesssim (2\pi)^{-1}$; besides, the spherical wave intensity should be noticeable compared to the "regular" OV beam. Because of both requirements, the ripple structure is best seen in the near field while on further propagation it becomes smoother due to smoothening wavefront and completely disappears in the far field since the spherical wave amplitude rapidly vanishes due to high divergence.

Another important observation relates to conditions under which, in small-angle diffraction regime, diffracted beams with a higher-order OV are generated. It turns out that in cases of non-unity diffraction order or non-unity topological charge of the embedded phase singularity, when, in accordance with Eq. (1), the CGH would produce a higher-order OV [11–13], this OV immediately decomposes into single-charged secondary OVs. This expected consequence of the asymmetric perturbation of the diffracted beam is seen very well in the 3$^{rd}$ row of Fig. 2. In fact, at any distance behind the CGH where the diffracted beams are well-formed and become available for observation, no higher-order OV can be obtained. Instead, arrays of single-charged OVs, "nested" within the diffracted beam, are formed. Visually, these arrays are analogous to those emerging upon astigmatic focusing of the higher-order LG beams [18]; however, detailed behavior of the spatial distribution and morphology parameters of the secondary OVs are quite different, which will be considered in the next Section.

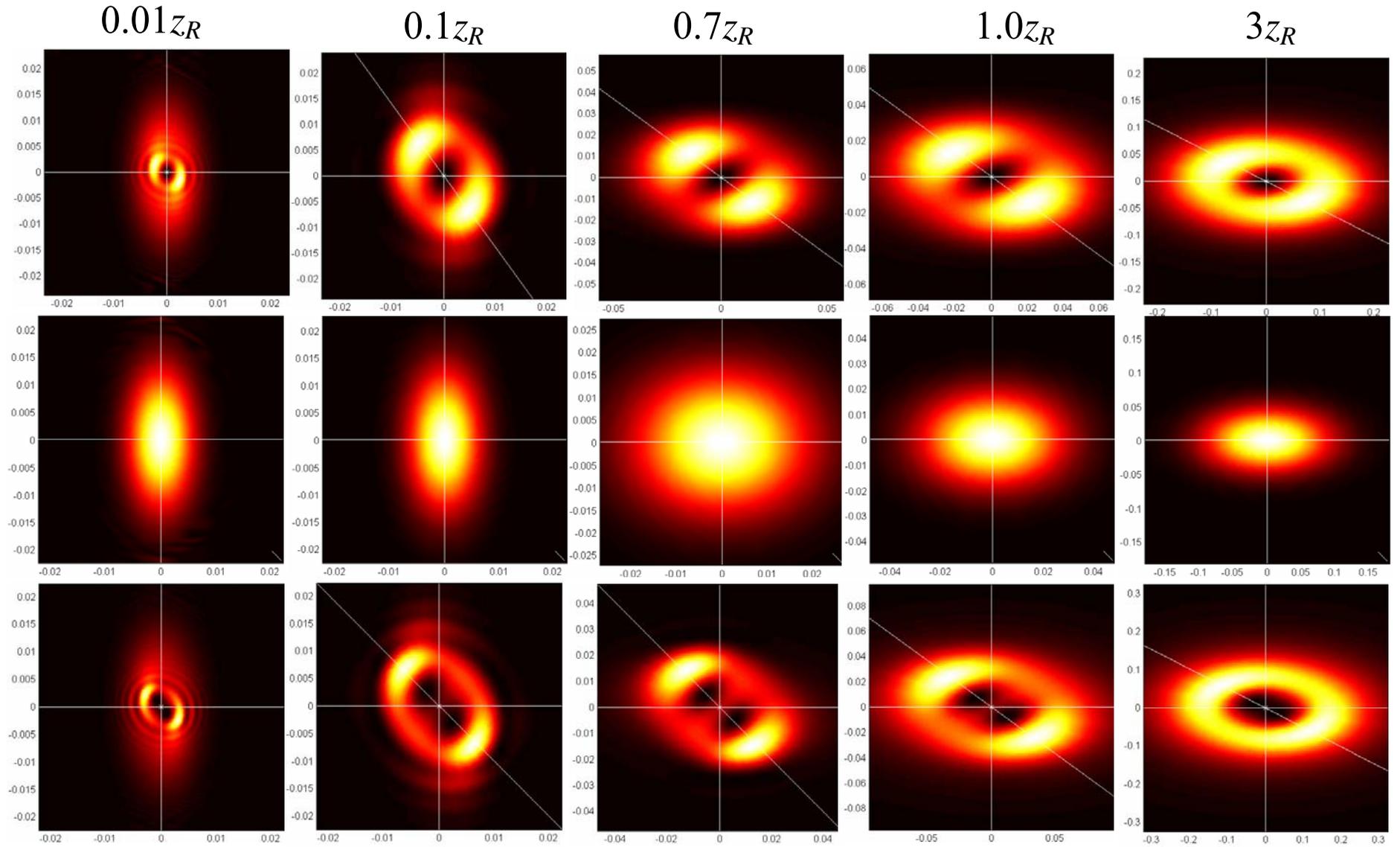

Fig. 2. Evolution of the transverse profile of the CGH-generated beam calculated by formula (12) at $\theta = 62°$ (squeezing coefficient (4) $\sigma = 2.13$), $\xi_s = \eta_s = 0$ for $l = 1$ (top row) and $l = 2$ (bottom row). For comparison, evolution of a Gaussian beam with the same initial squeezing is presented in the middle row. Propagation distances are marked above each column.

## 3. Characteristics of the OV arrays formed by the CGH when |l| > 1

In the current literature, the following characteristics of arrays of the single-charged OVs "nested" within a paraxial beam are commonly accepted: (i) their distribution, i.e. positions of the OV cores (amplitude zeros) in the transverse cross section [18,19,28], and (ii) morphology parameters, describing the field distribution near the individual OV cores [24–27]. Their definition is based on the fact that, for every cross section, in the nearest vicinity of an OV core, the complex amplitude distribution can be represented as

$$u_l(\xi,\eta,\zeta) \propto (g+f)(\xi-\xi_q) + i(g-f)(\eta-\eta_q) \tag{14}$$

where $g$ and $f$ are certain complex numbers, $\xi_q$ and $\eta_q$ are dimensionless Cartesian coordinates of the OV core normalized like Eq. (9); $g$, $f$, $\xi_q$ and $\eta_q$ generally depend on $\zeta$. The corresponding intensity distribution possesses rectangular symmetry and its constant-level contours are ellipses [27] (see Fig. 3 for examples). Usual characteristics of the OV morphology can be expressed via geometric parameters of these ellipses: angle of orientation $\theta_a$ and the ellipse form-factor (major to minor axes ratio $w_1/w_2$) as shown in Fig. 3.

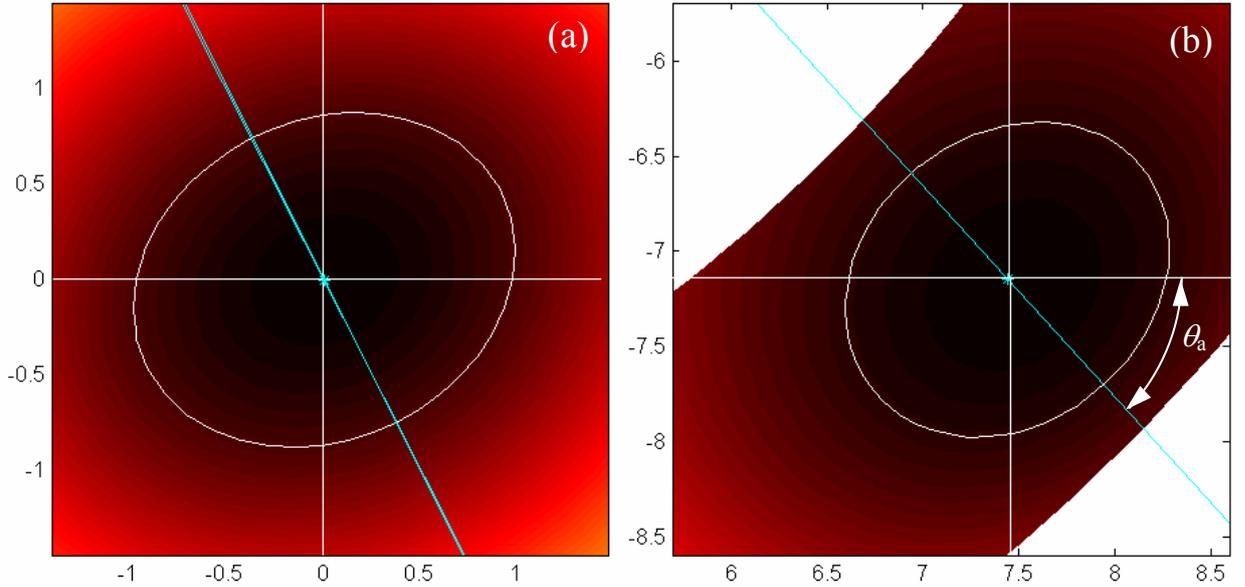

Fig. 3. Intensity pattern in the nearest vicinity of the secondary OV cores for the diffracted beam with $l = 3$, $\theta = 0.854$ rad ($\sigma = 1.5$) at the distance $\zeta = 0.35$ after the grating: (a) near the beam axis (axial OV); (b) near the off-axial OV. Ellipses of equal amplitude and the OV morphology parameters are shown.

In the numerical analysis, we directly calculate the beam intensity distribution near the expected OV core by means of formula (3) or (12) and afterwards, a contour of constant intensity is determined and fitted by an ellipse whose center, orientation and half-axes are found by using the least square approximation (see Fig. 3). The ellipse center is then identified with the OV core position; other parameters are used for the OV morphology analysis (Sec. 3.2).

*3.1. Positions of the OV cores within the beam cross section*

The high-order OV, expected when |$l$| > 1 in Eq. (1), can be treated as a "prototype" non-perturbed situation where all the secondary OVs are concentrated on the beam axis ($\zeta = 0$). With further propagation, the single-charged OVs separate and move away from the axis. In agreement with general considerations [1,3–5] and observations of other cases of the high-order OV perturbation [18,19], the total number of the secondary OVs equals to |$l$|. In accordance with the central symmetry of the transverse beam pattern, mentioned in Sec. 2, in any cross section the OVs are distributed symmetrically with respect to the beam axis. Due to this symmetry, if $l$ is an odd number, one of the single-charged OVs remains on the beam axis. All other OVs (and all the secondary OVs if $l$ is even) are situated in the opposite quadrants of the Gartesian frame. Qualitatively, their positions obey the simple rules formulated primarily for the case of astigmatic transformation of LG modes [18,19] (see Fig. 4 of Ref. [19]). As the beam "contracts" in certain transverse direction, the secondary OVs move as if they are "squeezed out" perpendicularly to the axis of the beam "compression", simultaneously experiencing certain additional deviation in agreement with handedness of the transverse energy circulation. In all examples of this paper the prototype beam possesses positive $l$ (counter-clockwise energy circulation when viewing against the beam propagation) and is squeezed along the $x$ axis, so the secondary OVs "slip out" along the $y$ axis and, additionally, displace into the $2^{nd}$ and $4^{th}$ quadrants of the Cartesian frame (sometimes much farther than in vertical direction, see Fig. 2, bottom row, and Fig. 4). For negative $l$, the beam patterns would differ by the mirror-like reflection with respect to the vertical axis. This symmetry enables one to consider only the OVs situated at $x \geq 0$, which is employed in Figs. 3 – 6.

All the formulated regularities of the secondary OVs' displacements are similar to the analogous properties of the secondary OVs formed under astigmatic transformation of a high-order LG mode [18]. However, the detailed picture of their distribution in the OV beam formed by the "fork" CGH is a bit different. For example, all the secondary OVs, generated after the astigmatic transformation of an LG mode, in every cross section lie on a single straight line intersecting the beam axis [18,19]. To check this property for the situation of this paper, one should note that for |$l$| $\leq 3$ all the OVs lie on a straight line because of clear geometric requirements, including the above mentioned central symmetry. For $l = 4$ there exists a pair of "inner" secondary OVs and a pair of "outer" ones; Fig. 4 shows those situated in the $4^{th}$ quadrant. One can see that, in contrast to the data of Refs. [18,19], their cores do not belong to a straight line: there are distinct angles between the solid lines connecting positions of the two OV cores and lines connecting the inner cores with the beam axis, at least for $\zeta = 0.35$ and $0.7$. These angles gradually vanish only when $\zeta \to 0$ and $\zeta \to \infty$.

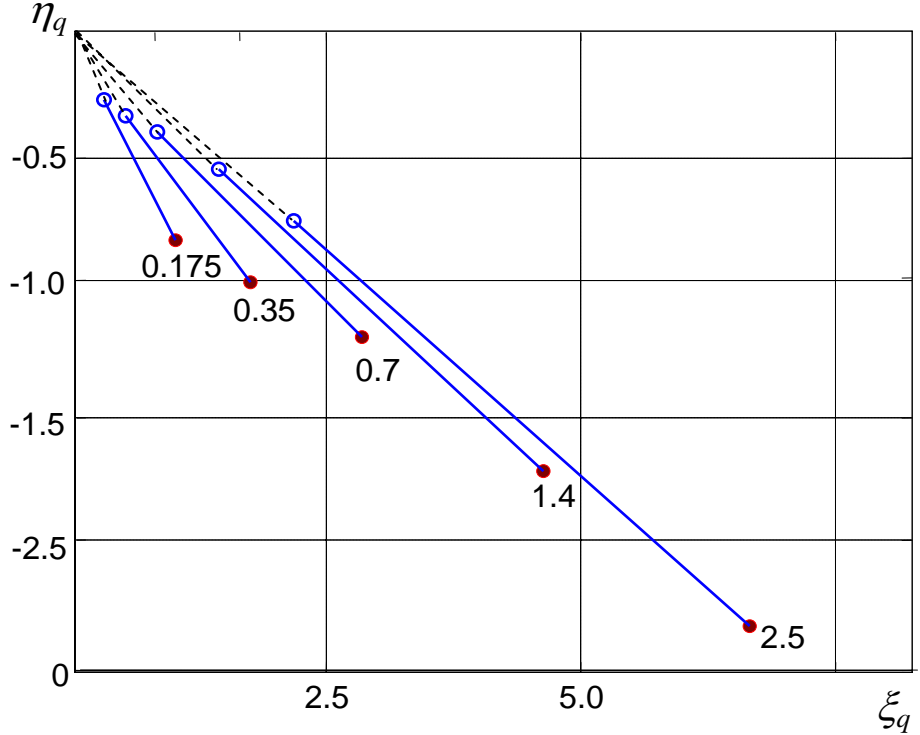

Fig. 4. Positions of the secondary OVs within the 4$^{th}$ quadrant of the Cartesian frame in the diffracted beam cross section for $l = 4$, $\theta = 1.23$ rad (70.5°): inner (open circles) and outer (filled circles) OVs of the same cross sections are connected by solid lines, propagation distances in units of $\zeta$ are indicated near each filled circle.

Another interesting issue related to the secondary OVs is quantitative description of their "moving away" from the axis during the diffracted beam propagation. Besides the general interest, it is important in the light of possible use of such beams for the generation of the OV arrays [18,28–31]. Corresponding results are presented in Figs. 5 – 7.

Fig. 5 and Appendix show that when solving this problem it is convenient to consider an intermediate situation where the output beam deformation described by the coefficient (4) is negligible but decomposition of the expected $l$-charged OV is already noticeable. In this situation, by using the known expansion

$$\cos\theta \approx 1 - \frac{1}{2}\theta^2 + \frac{3}{8}\theta^4 + \dots, \qquad (15)$$

that is correct when

$$\theta \ll 1, \qquad (16)$$

and assuming that near the OV cores

$$\frac{\rho^2}{2\zeta} = k\frac{r^2}{2z}\theta^2 \ll 1 \tag{17}$$

($\rho = \sqrt{\xi^2 + \eta^2}$), one can derive analytical estimates for the OV coordinates (A.16), (A.21). For the case without misalignment considered in this paper ($\xi_s = \eta_s = x_s = y_s = 0$ in (5), (8)) they give very simple expressions

$$\xi_q \approx -\text{sgn}(l)\eta_q = \pm\theta\sqrt{C_l\zeta}. \tag{18}$$

Coefficients $C_l$ for the most real situations where $l$ vary from 2 to 5 are presented in Table 1.

Table 1. Coefficients in expressions (18) and (20) for the OV positions

| $l$ | $C_l = C_{-l}$ |
|---|---|
| 1 | 0 |
| 2 | $\dfrac{1}{2}$ |
| 3 | $\dfrac{3}{2}$ |
| 4 | $\dfrac{3 \pm \sqrt{6}}{2} \approx \begin{cases} 2.725 \\ 0.275 \end{cases}$ |
| 5 | $\dfrac{5 \pm \sqrt{10}}{2} \approx \begin{cases} 4.081 \\ 0.919 \end{cases}$ |

As Fig. 5 witnesses, although for chosen values of $\theta$ the approximation leading to (18) seems rather rough (for example, when $\theta = 0.455$, two terms of (15) provide the accuracy limited by 10%), its quality is considerably high. It falls down with growing $|l|$, which is explained by the growing number of terms in expansion (15) necessary for the calculations. Note that in (18) as well as in other approximations that can be derived analogously (see Appendix), $\xi_q$ and $\eta_q$ are always mutually proportional and, what is more, their magnitudes are equal. This means that under conditions (15) – (17), unlike the more general case presented in Fig. 4, all the secondary OVs lie on a single straight line that coincides with one of the bisectors of the coordinate angles. The calculated points in Fig. 5 generally confirm this conclusion. Only at $l = 4$ and very close to the grating $\xi_q$ and $\eta_q$ may differ (filled and empty markers near curves 1 and 2 do not coincide).

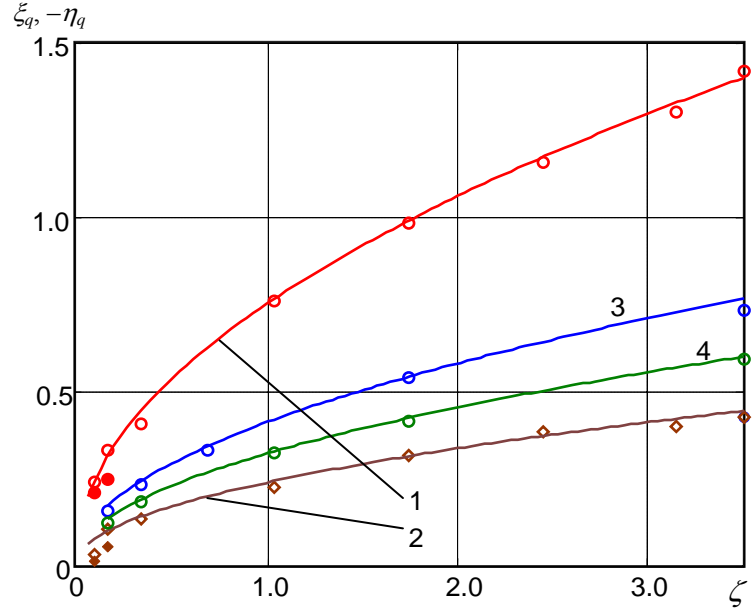

Fig. 5. Coordinates of the secondary OV cores vs propagation distance for moderate diffraction angles: numerical data (markers) and corresponding approximations (18) (curves); empty markers denote $\xi_q$ and coinciding $\xi_q$ and $\eta_q$, filled markers – $\eta_q$.
(1) and (2): $l = 4$, $\theta = 0.455$ rad (26°), circles – outer OVs, diamonds – inner OVs, approximation curves 1 and 2 differ by the sign in $C_4$;
(3) $l = 3$, $\theta = 0.336$ rad (19.2°), approximation (18) for $C_3$;
(4) $l = 2$, $\theta = 0.455$ rad, approximation (18) for $C_2$ (values $C_l$ see in Table 1).

Approximation (18) becomes insufficient if the output beam experiences strong deformation in the plane of diffraction (coefficient (4) essentially differs from the unity). In such cases simplifications of Eq. (12) are available only in the far field where

$$k\frac{r_a^2}{2z} \sim \zeta^{-1} \ll 1, \tag{19}$$

so the approximation is valid at arbitrary $\theta$ but in the far field only. In spite of different assumption, calculations of the OV positions carried out in the Appendix give the quite similar to (A.21) result (A.27), which in the case of perfect alignment reduces to

$$\xi_q \approx -\mathrm{sgn}(l)\frac{\eta_q}{\cos\theta} = \pm\tan\theta\sqrt{C_l\zeta}. \tag{20}$$

with the same coefficients $C_l$ of Table 1 that occur in (18). The data of Fig. 6 witness that this approximation qualitatively "works" even at relatively small $\zeta$; agreement with the exact results of numerical analysis is better for $\eta_q$, i.e. for the minor coordinates of the OV cores. For comparison, Fig. 6 also presents displacements of the secondary OVs obtained after the astigmatic telescopic transformation of the $LG_{0l}$ mode whose Gaussian envelope coincides with the incident Gaussian beam, i.e. for the beam with initial ($z = 0$) complex amplitude distribution

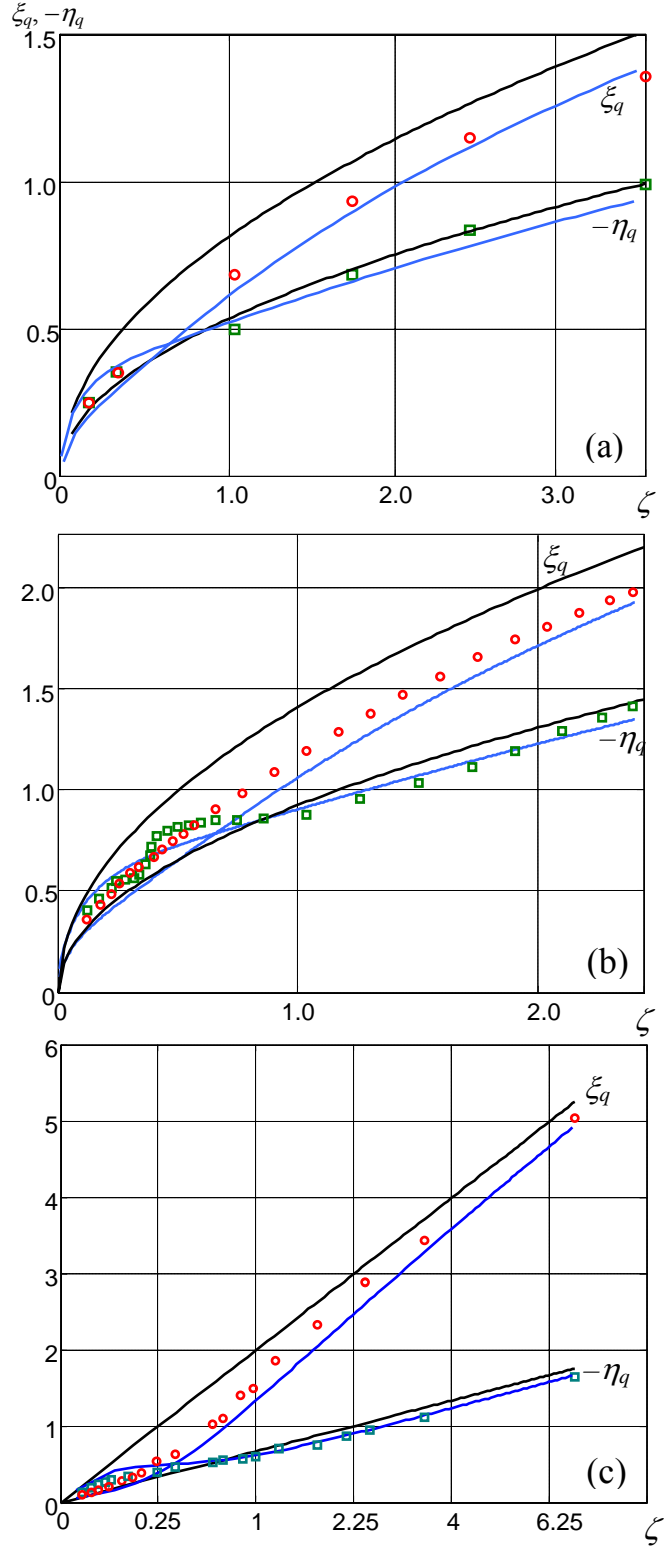

Fig. 6. Coordinates of the secondary OV cores vs propagation distance for high diffraction angles; markers denote numerical data (circles – $\xi_q$, squares – $\eta_q$), black curves – corresponding approximations (20), light curves describe positions of the OV cores for asymmetrically deformed LG modes with the same $l$ and squeezing coefficient $\sigma = (\cos\theta)^{-1}$ (initial distribution (21)).
(a) $l = 2$, (b) $l = 3$; in both (a) and (b) $\theta = 0.854$ rad $= 49°$, $\sigma = 1.5$;
(c) $l = 2$, $\theta = 1.23$ rad $= 70.5°$, $\sigma = 3.0$; note the square-law $\zeta$ scale that enables better resolution of calculated points in the region of small $\zeta$ and provides rectification of the black curves.

$$u_{0l}^{LG}(\xi,\eta) \propto \left(\frac{x}{b_x} + i\,\text{sgn}(l)\frac{y}{b}\right)^{|l|} \exp\left(-\frac{x^2}{2b_x^2} - \frac{y^2}{2b^2}\right) = [\sigma\xi + i\,\text{sgn}(l)\eta]^{|l|} \exp\left(-\frac{\sigma^2\xi^2 + \eta^2}{2}\right) \quad (21)$$

where $b_x = b/\sigma$ and $\sigma$ is the squeezing coefficient (4) corresponding to the diffraction angle (light curves were calculated by formula (22) of Ref. [18]). These curves provide quite reasonable approximation of the OV positions for the diffracted beams generated by the CGH. This is rather surprising, since the "overall" spatial characteristics of the CGH-produced OV beam essentially differ from those of the LG modes even in the small-angle diffraction limit [12,13].

In the region $\zeta < 0.5$, there appears a visually irregular oscillating component in the behavior of $\xi_q$, $\eta_q$ as functions of $\zeta$. This is especially noticeable by the distinct kink in the distribution of square markers in Fig. 6b ($\eta_q(\zeta)$ dependence), and can be attributed to the influence of the ripple structure discussed in the Section 2 (see 1st and 2nd panels in the bottom row of Fig. 3).

Like in case of a deformed high-order LG mode [18], it is interesting to inspect the relative separation of the secondary OVs with respect to the current beam profile, which may be useful in the context of creation of the OV arrays. Since the diffracted beam intensity distribution is represented by complicated functions without explicit analytical expression, it is convenient to characterize the beam profile by means of the second intensity moments [32–35] which form the symmetric positive definite matrix

$$M_{11} = \left[\int |u_{le}(\xi,\eta)|^2 \, d\xi d\eta\right]^{-1} \int \begin{pmatrix} \xi^2 & \xi\eta \\ \xi\eta & \eta^2 \end{pmatrix} |u_{le}(\xi,\eta)|^2 \, d\xi d\eta. \quad (22)$$

The integrals should be calculated over the whole cross section. The intensity moments (22) are known as general characteristics of the beam transverse shape. However, for the CGH-produced OV beams satisfying Eq. (8) or (12), the second integral in Eq. (22) diverges. This occurs because the intensity distribution $|u_{le}(\xi,\eta)|^2$ falls down rather slowly at the beam periphery [12] so the role of the integrand "tails" at $\xi \to \infty$, $\eta \to \infty$ is overestimated. Since we are mainly interested in the characteristics of the visible beam profile, we artificially restrict the integration domain by the square boundary centered at the beam axis and determined by requirement that the boundary intensity does not exceed 1% of the beam intensity maximum. Then we evaluate Eq. (22) numerically and assume the amount $\sqrt{\text{Sp}(M_{11})}$ (Sp is the symbol of the matrix trace) to be a measure of the beam transverse size [35]. The relative separation of the OV cores can be estimated as $\xi_q/\sqrt{\text{Sp}(M_{11})}$. Corresponding results are presented in Fig. 7 and show that, indeed, there exists a distance at which the secondary OVs are separated most expressively (maximums of the black curves). This distance lies in the near field and depends on the diffraction angle (or, rather, on the squeezing coefficient (4)) so that it diminishes when the initial beam deformation becomes stronger.

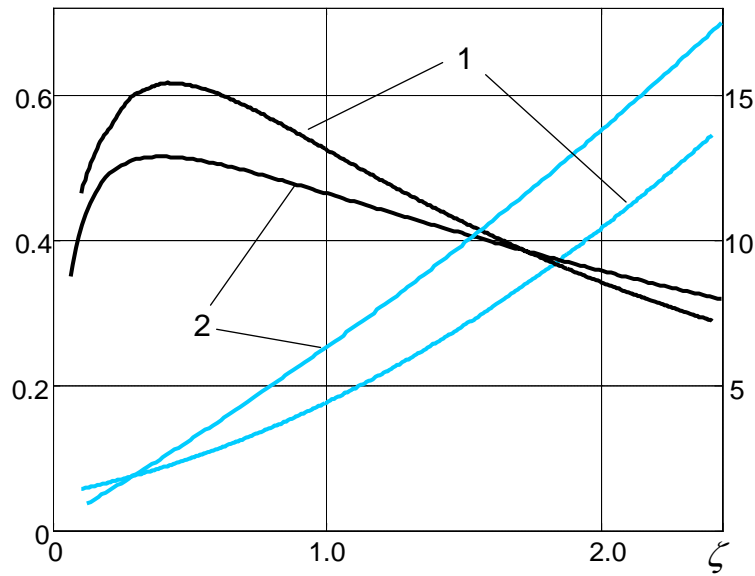

Fig. 7. Evolution of the beam transverse size $\sqrt{\mathrm{Sp}(M_{11})}$ (light curves, right vertical scale) and of the relative separation of the OV cores $\xi_q/\sqrt{\mathrm{Sp}(M_{11})}$ (black curves, left vertical scale): (1) $l = 3$, $\theta = 0.854$ rad (beam of Fig. 6b); (2) $l = 2$, $\theta = 1.23$ rad (beam of Fig. 6c).

*3.2. Morphology parameters of the secondary OVs*

As is shown in the Appendix (Sec. A.3), analytical approximations for the morphology parameters appear to be useless for the investigation of their evolution. Under conditions (15) – (17) they predict that all secondary OVs are isotropic, just like the "prototype" multicharged OVs generated by the CGH when $\cos\theta \approx 1$ [11–13]; for the far-field condition of Eq. (19), they give only the constant asymptotic values $\theta_a = \pi/2$ and $w_1/w_2 = \sigma$ corresponding to $\zeta \to \infty$. That is why behavior of the secondary OV morphology was studied numerically on an example of the diffracted beam with $l = 3$, $\theta = 0.854$ rad (squeezing coefficient (4) is $\sigma = 1.5$). The results are presented in Figs. 3 and 8; they are considered in comparison with the morphology parameters of the LG$_{03}$ beam experiencing the same transverse deformation (with initial complex amplitude distribution (21)).

The first feature that articulately differs the studied CGH-produced OV beam from its LG analog is that the morphology of separate individual secondary OVs is not the same. It is qualitatively visible even in Fig. 3 that allows to compare the intensity patterns near the axial OV and near the OV displaced from the axis (the displacement itself was discussed above, see Fig. 6b).

The quantitative characterization of different morphologies is provided by Fig. 8. At moderate distances $\zeta$ after the CGH, orientation angle $\theta_a$ of the off-axial OV grows a bit faster than orientation angle of the axial one (Fig. 8a); herewith, both OVs change their orientations more rapidly than in case of the corresponding deformed LG beam (dashed curve). The OV form-factor behavior (Fig. 8b) also shows the "overall" grows with increasing $\zeta$ but now the form-factor of the axial OV "leads" in the general tendency to the asymptotic value $w_1/w_2 = \sigma$. Note that the very variability of the secondary OV form-factors in these conditions is a peculiarity of the CGH-produced beams because for the LG analog this form-factor is constant (dashed line in Fig. 8b).

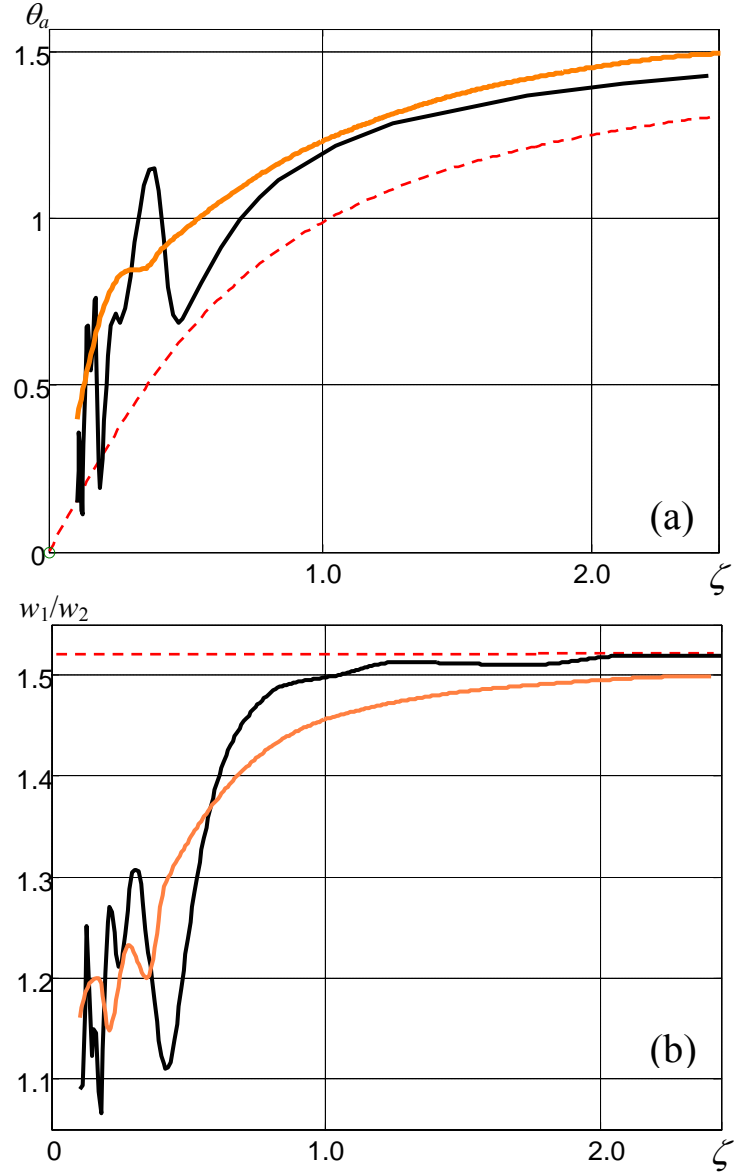

Fig. 8. Evolution of the morphology parameters of the secondary OVs for the diffracted beam with $l = 3$, $\theta = 0.854$ rad, $\sigma = 1.5$ (beam of Fig. 6b): (a) orientation angle $\theta_a$, (b) form-factor $w_1/w_2$. Black curves: axial OV, light curves: off-axial OV, dashed lines: OV of the asymmetrically deformed LG modes with the same $l$ and squeezing coefficient $\sigma$ (initial distribution (21)).

But the most impressive distinction from the LG analog is the irregular oscillations seen in behavior of all the presented in Fig. 8 morphology parameters at $\zeta \lesssim 0.5$. This clearly corresponds to analogous performance of the OV core coordinates in Fig. 6b and can be related to the ripple structure discussed in Sec. 2 (see 1st and 2nd columns of Fig. 2). Like the ripples, these oscillations owe to the influence of the interference with the divergent spherical wave originating from the bifurcation point. Its stronger action on the parameters of the axial secondary OV can be explained by the symmetry of corresponding perturbation. Circular interference pattern more strongly affects the "tilted" ellipse of equal amplitude centered at the beam axis, than the similarly oriented ellipse of the off-axial OV, because the latter is elongated approximately in the same azimuthal direction as the ripple fringes nearby (see, for example, Fig. 3a, b).

## 4. Conclusion

In this paper, the analysis has been performed of spatial properties of the OV beams obtained with the help of the "fork" hologram in conditions when, due to high diffraction angle, requirement $\cos\theta \approx 1$, accepted in previous works [11–13], is violated. In this situation, the incident beam diffraction is coupled not only with the OV formation but also with the beam squeezing in the diffraction plane. The latter transformation is equivalent to the astigmatic telescopic transformation and evokes the same main consequences that accompany any symmetry breakdown of an initially circular OV beam [18–21]. At the moment of transformation, the beam circular symmetry reduces to the rectangular one; then, upon propagation, the rectangular symmetry is also destroyed and the beam profile experiences rotation of the same sense as the transverse energy circulation in the initial (unperturbed) beam. The rotation asymptotically stops and the beam rectangular symmetry is restored only in the far field. The overall behavior of the CGH-produced OV beam is similar to the behavior of the astigmatically transformed LG mode [18,19] but differs from it in some quantitative details. The most impressive differences occur in the near field due to interference with the divergent spherical wave originating from the bifurcation point [12,22], which is manifested in the ripple structure imposed over the beam amplitude and phase distributions.

The diffracted beam evolution looks more complicated when the topological charge of the phase singularity embedded in the CGH $|m| > 1$ or/and the diffraction order $|n| > 1$ (usual conditions for the high-order OV generation). In this case, simultaneously with the diffracted beam formation, this hypothetical high-order OV is decomposed into a set of single charged ones. In contrast to the case of astigmatically transformed LG beam, these secondary OVs are not always distributed along a single straight line; however, evolution of their positions in the propagating beam cross section can be described by corresponding LG-beam-related formulas with reasonable accuracy (Fig. 6). In the wide range of diffraction angles $\theta$ and propagation distances $z$, the secondary OV positions can be characterized by simple asymptotic formulas (18), (20) (see Fig. 5) which are also applicable to the secondary OVs emerging in case of astigmatic transformation of high-order LG beams.

The secondary OVs are in general anisotropic. The behavior of their morphology parameters (orientation in the transverse cross section and the constant-intensity ellipse form factor) with the beam evolution has been studied numerically. Compared to the astigmatically transformed LG modes, they show pronounced dissimilarities (Fig. 8). In particular, in the near field, the morphology of secondary OVs is noticeably affected by the ripple structure, and this effect is stronger for the OV situated at the beam axis; besides, over the whole beam evolution, except the far field, the morphologies of different individual OVs nested within the same OV beam are no longer identical. In the far field, the morphology characteristics approach the asymptotic values common with those of secondary OVs originating from the LG beams experiencing the same squeezing.

The results of our analysis will be useful for creation of the OV beams with necessary properties. In particular, the effect of multiple OV formation in conditions when $|l| > 1$ (see Eq. (1)) camn be used for the formation of the OV arrays [28–31]; the existence of the maximum relative separation of the secondary OV cores (Fig. 7) can be a guideline in the search of corresponding transformation arrangement.

On the other hand, the analysis presented can serve to more exactly specify the range of validity for approximation $\cos\theta \approx 1$ used in the previous works [11–15], whose important feature is absence of the high-order OV decomposition in the CGH-generated beams. This can be appropriate if the real secondary OV deviation is small compared to the beam transverse size. For example, if the CGH with 16 grooves per millimeter is used (the case of Refs. [12,13]), according to (2), the diffraction angle amounts to $\theta = 10^{-2} n$ rad. Due to Eq. (18), this means that for the $n$-order diffracted beam, separation of secondary OVs roughly equals to $n$ % of the incident beam size, and a noticeably lesser part of the current beam size (in conditions of Refs. [12,13], $x_q = y_q = b\xi_q = b\eta_q \approx n$ micrometers). Usually, such a small effect can readily be masked by the noise and/or by the limited resolution of the image analyzing setup, so the secondary OVs visually "combine" into a single high-order OV.

**Appendix**

*A.1. Analytical study of the secondary OVs' separation in case $\theta \ll 1$*

To clarify the idea of the analytical estimate, let us start with considering the two-term approximation of (15), keeping only one $\theta$-dependent term proportional to $\theta^2$. Then

$$\left(\xi - \xi_a \cos\theta\right)^2 \approx \left(\xi - \xi_a\right)^2 + \xi_a\left(\xi - \xi_a\right)\theta^2 \tag{A.1}$$

Under additional condition (17), expansion

$$\exp(t) \approx 1 + t + \frac{1}{2}t^2 + ... \tag{A.2}$$

($t \ll 1$) can be used; keeping only two terms of this expansion, Eq. (12) can be represented in the form

$$u_{le}(\xi,\eta,\zeta) = \left[u_{le}^0(\xi,\eta,\zeta) + \Delta u_{le}(\xi,\eta,\zeta)\right], \tag{A.3}$$

where

$$u_{le}^0(\xi,\eta,\zeta) = \frac{1}{2\pi i\zeta}\int u_a(\xi_a,\eta_a)e^{il\phi}\exp\left\{\frac{i}{2\zeta}\left[\left(\xi-\xi_a\right)^2+\left(\eta-\eta_a\right)^2\right]\right\}d\xi_a d\eta_a \tag{A.4}$$

is a function describing the diffracted beam in approximation $\cos\theta \approx 1$, and

$$\Delta u_{le}(\xi,\eta,\zeta)$$
$$= \frac{1}{2\pi i\zeta}\int u_a(\xi_a,\xi_a)e^{il\phi}\exp\left\{\frac{i}{2\zeta}\left[\left(\xi-\xi_a\right)^2+\left(\eta-\eta_a\right)^2\right]\right\}\frac{i}{2\zeta}\theta^2\left(\xi-\xi_a\right)\xi_a\, d\xi_a d\eta_a$$
$$= \frac{i}{2\zeta}\theta^2\left(-i\zeta\xi\frac{\partial}{\partial\xi} - i\zeta + \zeta^2\frac{\partial^2}{\partial\xi^2}\right)u_{le}^0(\xi,\eta,\zeta) \tag{A.5}$$

(the rule of differentiating an integral with respect to a parameter is used). Therefore, Eq. (A.3) reduces to

$$u_{le}(\xi,\eta,\zeta) = \left[1 + \frac{1}{2}\theta^2\left(\xi\frac{\partial}{\partial\xi} + 1 + i\zeta\frac{\partial^2}{\partial\xi^2}\right)\right]u_{le}^0(\xi,\eta,\zeta). \tag{A.6}$$

For the case of incident Gaussian beam when $u_a(\xi_a,\eta_a)$ is defined by Eq. (13), behavior of function (A.4) is well known [12,13,22], and in the near vicinity of the OV cores it is described by formulae (18), (21) and (15), (16) of Ref. [13]. They read

$$u_{le}^0(\xi,\eta,\zeta) = D(\zeta)\exp\left[\frac{i}{2\zeta}\left(\xi^2+\eta^2\right)\right]\left[\Sigma(\xi,\eta)\right]^{|l|}, \tag{A.7}$$

where $D(\zeta)$ does not depend on $\xi$ and $\eta$,

$$\Sigma(\xi,\eta) \equiv \Sigma = \xi - \xi_V + i\,\text{sgn}(l)(\eta - \eta_V). \tag{A.8}$$

Here

$$\xi_V = \text{sgn}(l)\zeta\eta_s, \quad \eta_V = -\text{sgn}(l)\zeta\xi_s \tag{A.9}$$

are non-perturbed coordinates of the high-order OV core calculated in the assumption $\cos\theta \approx 1$ [13] ($\xi_s$, $\eta_s$ are dimensionless coordinates of the incident beam displacement).

To find positions of the secondary OV cores, one should equate function (A.6) to zero. In accordance with (A.7) and (A.8), near the beam axis $u_{le}^0(\xi,\eta,\zeta) \propto \Sigma^{|l|}$. We also make a supposition, that will be justified later, that near the sought zeros of function (A.6)

$$\Sigma \sim \theta\sqrt{\zeta}. \tag{A.10}$$

This means that near the OV cores

$$u_{le}^0(\xi,\eta,\zeta) \sim \theta^{|l|}, \quad \frac{\partial^n u_{le}^0(\xi,\eta,\zeta)}{\partial \xi^n} \sim \theta^{|l|-n}. \tag{A.11}$$

Therefore, in (A.6) the terms with highest derivatives dominate and other terms in parentheses can be omitted, so in the vicinity of an OV core

$$u_{le}(\xi,\eta,\zeta) = \left(1 + \frac{1}{2}\theta^2\zeta\frac{\partial^2}{\partial\xi^2}\right)u_{le}^0(\xi,\eta,\zeta). \tag{A.12}$$

After substitution of (A.7) and omitting inessential factor $D(\zeta)$ this expression reduces to

$$u_{le}(\xi,\eta,\zeta) \propto \exp\left[\frac{i}{2\zeta}(\xi^2+\eta^2)\right]P[\Sigma(\xi,\eta)] \tag{A.13}$$

where

$$P(\Sigma) = \left[\Sigma^{|l|} + i\frac{1}{2}\zeta\theta^2\left(\frac{i}{\zeta}\Sigma^{|l|} - \frac{1}{\zeta^2}\xi^2\Sigma^{|l|} + 2\frac{i}{\zeta}\xi|l|\Sigma^{|l|-1} + |l|(|l|-1)\Sigma^{|l|-2}\right)\right]$$

$$\approx \left[\Sigma^2 + \frac{i}{2}\zeta\theta^2|l|(|l|-1)\right]\Sigma^{|l|-2} \tag{A.14}$$

(first three terms in parentheses appear due to differentiating the exponential pre-factor in (A.7); because of (A.10), they are negligible and are thus discarded in the second line). Hence, the equation for the OV positions just follows

$$\left[\Sigma^2(\xi,\eta) + \frac{i}{2}\zeta\theta^2|l|(|l|-1)\right]\Sigma^{|l|-2}(\xi,\eta) = 0. \tag{A.15}$$

Its solutions are

$$\xi - \xi_V = -\text{sgn}(l)(\eta - \eta_V) = \pm\frac{\theta}{2}\sqrt{|l|(|l|-1)\zeta}, \tag{A.16}$$

$$\xi - \xi_V = \eta - \eta_V = 0. \tag{A.17}$$

Solutions (A.16) exist for any $|l| > 1$ and describe single-charged OVs. Solution (A.17) appears if $|l| > 2$ and is physically meaningful only at $|l| = 3$ when it describes the single-charged OV whose position coincides with the "unperturbed" position of the "prototype" high-order OV. In other situations it corresponds to "nonphysical" $(|l| - 2)$-order OV that does not exist in reality. So, the above reasoning enables us to determine positions of at most three OVs while, in fact, there always exist exactly $|l|$ first-order secondary OVs. "Missing" OVs can be found by the analogous procedure, but higher degrees in expansions (15), (A.1) and (A.2) should be taken into account. For example, if we employ the same approximations but with accuracy of $\theta^4$ and again apply conditions (A.10), (A.11), instead of Eq. (A.12) we will have

$$u_{le}(\xi,\eta,\zeta) = \left(1 + \frac{i}{2}\theta^2\zeta\frac{\partial^2}{\partial\xi^2} - \frac{1}{8}\theta^4\zeta^2\frac{\partial^4}{\partial\xi^4}\right)u_{le}^0(\xi,\eta,\zeta), \tag{A.18}$$

the complex amplitude distribution can still be expressed in the form (A.13) but instead of (A.14) and (A.15) we get

$$P(\Sigma) = \left[\Sigma^4 + \frac{i}{2}\theta^2\zeta|l|(|l|-1)\Sigma^2 - \frac{1}{8}\theta^4\zeta^2|l|(|l|-1)(|l|-2)(|l|-3)\right]\Sigma^{|l|-2} \tag{A.19}$$

and

$$\left[\Sigma^4(\xi,\eta) + \frac{i}{2}\theta^2\zeta|l|(|l|-1)\Sigma^2(\xi,\eta) - \frac{1}{8}\theta^4\zeta^2|l|(|l|-1)(|l|-2)(|l|-3)\right]\Sigma^{|l|-2}(\xi,\eta) = 0. \tag{A.20}$$

This equation has already four non-zero solutions

$$\xi - \xi_V = -\text{sgn}(l)(\eta - \eta_V) = \pm\theta\left\{\frac{1}{8}\zeta|l|(|l|-1)\left[1 \pm \sqrt{1 - \frac{2(|l|-2)(|l|-3)}{|l|(|l|-1)}}\right]\right\}^{1/2} \tag{A.21}$$

that describe positions of the secondary OVs when $|l| = 4$ and 5; for $|l| > 5$ the "non-perturbed" solution (A.17) is $(|l| - 4)$-fold and correspond to the nonphysical $(|l| - 4)$-order OV which can be "decomposed" with allowance for additional terms in expansions (15), (A.1) and (A.2), and so on.

### A.2. Analytical approximation for arbitrary $\theta$

For large $\theta$ simplification of the integral (12) is available under condition (19), i.e. in the far field. Then, in the integrand of Eq. (12) the exponent can be transformed as follows:

$$\exp\left\{\frac{i}{2\zeta}\left[(\xi - \xi_a\cos\theta)^2 + (\eta - \eta_a)^2\right]\right\}$$

$$\approx \exp\left[\frac{i}{2\zeta}(\xi^2 + \eta^2)\right]\exp\left[-\frac{i}{\zeta}(\xi_a\xi\cos\theta + \eta_a\eta)\right]$$

$$\times \left[ 1 + \frac{i}{2\zeta}\left(\xi_a^2 \cos^2\theta^2 + \eta_a^2\right) - \frac{1}{8\zeta^2}\left(\xi_a^2 \cos^2\theta^2 + \eta_a^2\right)^2 + ... \right] \quad (A.22)$$

and function (12) can be represented as

$$u_{le}(\xi,\eta,\zeta) = \frac{1}{2\pi i\zeta} \cdot \exp\left[\frac{i}{2\zeta}\left(\xi^2 + \eta^2\right)\right]$$

$$\times \left[ 1 - \frac{i}{2}\zeta\left(\frac{\partial^2}{\partial \xi^2} + \frac{\partial^2}{\partial \eta^2}\right) - \frac{1}{8}\zeta^2\left(\frac{\partial^4}{\partial \xi^4} + 2\frac{\partial^2}{\partial \xi^2}\frac{\partial^2}{\partial \eta^2} + \frac{\partial^4}{\partial \eta^4}\right) + ... \right] F(\xi\cos\theta,\eta) \quad (A.23)$$

where

$$F(\xi,\eta) = \int u_a(\xi_a,\eta_a) e^{il\phi} \exp\left[-\frac{i}{\zeta}(\xi_a\xi + \eta_a\eta)\right] d\xi_a d\eta_a .$$

To find explicit representation of this function, note that the far-field form of function (A.4) is

$$u_{le}^0(\xi,\eta,\zeta)_{\zeta \to \infty} \propto \exp\left[\frac{i}{2\zeta}(\xi^2 + \eta^2)\right] F(\xi,\eta) ;$$

on the other hand, the equivalent expression can be derived from (A.7):

$$u_{le}^0(\xi,\eta,\zeta)_{\zeta \to \infty} \propto \exp\left[\frac{i}{2\zeta}(\xi^2 + \eta^2)\right] \Sigma^{|l|}(\xi,\eta) .$$

As a result, we can accept

$$F(\xi\cos\theta,\eta) = \left[\Sigma(\xi\cos\theta,\eta)\right]^{|l|} . \quad (A.24)$$

Then, after substituting (A.24) into (A.23) and performing the necessary transformations, we obtain the complex amplitude representation in the form

$$u_{le}(\xi,\eta,\zeta) \propto \exp\left[\frac{i}{2\zeta}(\xi^2 + \eta^2)\right] P\left[\Sigma(\xi\cos\theta,\eta)\right] \quad (A.25)$$

where $P(\Sigma)$ is given by Eq. (A.19). Expression (A.25) differs from (A.13) only by the first argument of $\Sigma$. Positions of the secondary OVs follow from the requirement $u_{le}(\xi,\eta,\zeta) = 0$ and are determined by equation

$$\left[\Sigma^4(\xi\cos\theta,\eta) + \frac{i}{2}\zeta\sin^2\theta|l|(|l|-1)\Sigma^2(\xi\cos\theta,\eta) - \frac{1}{8}\zeta^2\sin^4\theta|l|(|l|-1)(|l|-2)(|l|-3)\right]$$

$$\times \left[\Sigma(\xi\cos\theta,\eta)\right]^{|l|-4} = 0 . \quad (A.26)$$

It appears to be quite similar to (A.20), and its solutions with allowance for (A.8) are similar to (A.21), (A.17):

$$\xi - \frac{\xi_V}{\cos\theta} = -\text{sgn}(l)\frac{(\eta - \eta_V)}{\cos\theta} = \pm\tan\theta\left\{\frac{1}{8}\zeta|l|(|l|-1)\left[1\pm\sqrt{1-\frac{2(|l|-2)(|l|-3)}{|l|(|l|-1)}}\right]\right\}^{1/2}, \quad (A.27)$$

$$\xi - \frac{\xi_V}{\cos\theta} = \eta - \eta_V = 0. \quad (A.28)$$

Note that Eqs. (A.27), (A.28) agree with the fact that in the far field the beam pattern stretches in the *x* direction proportionally to the squeezing coefficient (4).

Like in case of Eq. (A.20), these solutions describe a limited number of separate secondary OVs (at maximum, five), which is connected to the fact that in (A.22) we took only three terms of the exponent expansion (A.2). For |*l*| > 5, more complete set of the secondary OV positions can be calculated if additional terms in expansion (A.22) are taken into account.

*A.3. Morphology of the secondary OVs*

Approximate description of the morphology of the secondary OVs can be obtained directly from the explicit formulas (A.13) and (A.25) for the complex amplitude in close vicinity of the OV cores. Since the exponential prefactor does not affect the intensity distribution, the morphology parameters are fully determined by the polynomial term $P(\Sigma)$. Obviously, its constant-level contours coincide with those of $\Sigma$. This circumstance facilitates the OV morphology analysis in the approximation considered and, simultaneously, restricts its information value by rather trivial results. In case $\theta \ll 1$, corresponding to Sec. A.1, the discussed contours are circles, which means that all the secondary vortices are isotropic. In case of arbitrary $\theta$ and in the far field (Sec. A.2), the constant level contours of $\Sigma(\xi\cos\theta, \eta) = \xi\cos\theta - \xi_V + i\,\text{sgn}(l)(\eta - \eta_V)$ are horizontally elongated ellipses (orientation angle $\theta_a = \pi/2$) with form-factor $w_1/w_2 = \sigma$ (see Eq. (4)).